\documentclass[
 reprint,            
 superscriptaddress, 
 amsmath,amssymb,    
 aps,                
 prl,                
 floatfix            
]{revtex4-2}

\usepackage{graphicx}   
\usepackage{dcolumn}    
\usepackage{bm}         
\usepackage{hyperref}   
\usepackage{physics}    
\usepackage{xcolor}
\usepackage[normalem]{ulem}
\usepackage{soul}

\usepackage{hyperref}

\begin{document}

\title{Observing relativistic trajectories of single photons}

\author{Sayantan Das}
\email{syntds@gmail.com}
\affiliation{School of Mathematics and Physics, The University of Queensland, Brisbane, Queensland 4072, Australia}

\author{Daniel S. Dahl}
\affiliation{School of Physics, The University of Sydney, Sydney, NSW 2006, Australia}
\author{Daniel Peace}
\affiliation{School of Mathematics and Physics, The University of Queensland, Brisbane, Queensland 4072, Australia}
\author{Marcelo P. Almeida}
\affiliation{School of Mathematics and Physics, The University of Queensland, Brisbane, Queensland 4072, Australia}
\author{Abhishek Roy}
\affiliation{School of Mathematics and Physics, The University of Queensland, Brisbane, Queensland 4072, Australia}
\affiliation{Department of Physics and Astronomy, Macquarie University, Sydney, NSW 2113, Australia}
\author{Markus Rambach}
\affiliation{School of Mathematics and Physics, The University of Queensland, Brisbane, Queensland 4072, Australia}
\author{Andrew G. White}
\affiliation{School of Mathematics and Physics, The University of Queensland, Brisbane, Queensland 4072, Australia}
\author{Timothy C. Ralph}
\affiliation{School of Mathematics and Physics, The University of Queensland, Brisbane, Queensland 4072, Australia}
\author{Jacquiline Romero}
\affiliation{School of Mathematics and Physics, The University of Queensland, Brisbane, Queensland 4072, Australia}

\date{August 27, 2026}

\begin{abstract}

\noindent While the standard interpretation of quantum mechanics does not assign definite trajectories to particles, the Bohmian interpretation does. Only recently has an operational method for reconciling Bohmian mechanics with relativity been proposed. Here, we experimentally reconstruct relativistic Bohmian trajectories of a single photon in a Michelson-Sagnac interferometer, where counter-propagating probability amplitudes interfere head-on at the speed of light. As predicted by the relativistic Bohmian theory, we observe subluminal and superluminal features of the Bohmian trajectories of the photon traversing through the fringes. Our work provides experimental access to relativistic Bohmian mechanics and enables exploration of its unusual and counterintuitive properties.

\end{abstract}

\maketitle

\noindent Classical mechanics describes the preparation, evolution, and measurement of a particle's state in a deterministic manner. Given the position and velocity of a particle at a particular instant, classical dynamics uniquely determines its position and velocity at all future and past times. Quantum mechanics departs significantly from this classical picture. In the standard formulation of quantum mechanics, the notion of particles following definite trajectories ceases to exist, and physical systems are instead described by wavefunctions that yield only probabilistic predictions for measurement outcomes. Standard quantum mechanics accurately predicts experimental outcomes but says nothing of the nature of the quantum systems the theory describes, hence the variety of interpretations that attempt to draw meaning from the mathematical formalism. Among them, the de Broglie-Bohm interpretation or Bohmian mechanics \cite{PhysRev.85.166, PhysRev.85.180} is a nonlocal hidden variable interpretation of quantum theory, in which quantum particles follow deterministic trajectories, referred to as Bohmian trajectories. Different interpretations naturally motivate to investigate different quantities. In Bohmian mechanics, the notion of particle trajectories motivates the measurement of the particle velocities at all spacetime points, thus reconstructing the trajectories from these measurements.\\

\noindent An experimenter cannot observe a single Bohmian trajectory in the laboratory owing to Heisenberg's uncertainty principle prohibiting the simultaneous precise measurement of the particle's position and momentum. It has been shown by Wiseman in \cite{wiseman_grounding_2007} that the Bohmian trajectories of an ensemble of particles prepared in the same state can be reconstructed by weak-value measurements \cite{aharonov1988result, duck1989sense}. Reconstruction of such trajectories requires an experimenter to weakly measure the momentum of the particle and subsequently perform a strong measurement of its position. Repeating this procedure for such a post-selected ensemble of particles determines the average value of the particle's momentum known as the weak value. If we assume that all particles passing through the same point are following the same deterministic trajectory, then this weak value of momentum of the particle at a particular position is interpreted as proportional to the particle velocity at that position. Obtaining such weak-value-based Bohmian velocity at different positions allows the mapping of the Bohmian trajectories. This idea of using weak values to reconstruct a set of trajectories has been implemented in experiments which measured the transverse component of the velocity of single photons \cite{kocsis2011observing} and entangled photons \cite{mahler2016experimental} in a double-slit interferometer setup. The photons are slow in the transverse direction of propagation, making the measurement of the trajectories in this way non-relativistic.  \\

\noindent Bohmian mechanics had struggled to be consistent with the theory of special relativity, another cornerstone of classical physics. Foo et. al. \cite{foo2022relativistic} formulated a relativistic version of the weak-value measurement formalism which could be used to reconstruct the relativistic Bohmian trajectories of a single photon. Both the relativistic and non-relativistic frameworks are operational, in that the trajectories are based on the preparation and measurement of the system observables. In contrast to the non-relativistic regime, obtaining the Bohmian velocity field in the relativistic regime requires weakly measuring both the momentum and energy of the particle and a subsequent strong measurement on the position. The weak measurement of momentum is carried out in the non-relativistic experiments \cite{kocsis2011observing, mahler2016experimental} by placing a birefringent crystal in the path of the photon which weakly couples the transverse momentum of the photon with the polarisation. \\

\noindent In this work, we propose and implement a protocol to reconstruct relativistic Bohmian trajectories of a single photon interfering in a Michelson-Sagnac interferometer.
We show that, for our experimental conditions, Foo et. al.'s protocol \cite{foo2022relativistic} can be reduced to measuring the weak value of the photon's momentum as a function of its position in the propagation direction. Measuring this weak value along the photon's propagation direction, with the photon wavefunctions interfering head-on, manifests the relativistic nature of the measured trajectories. We perform the experiment with single photons from an attenuated continuous wave (CW) laser and plot the resulting relativistic Bohmian trajectories. The behaviour of our observed photon trajectories is radically different from the standard interpretation. The photons follow a single path through the interferometer but alternate between sub- and superluminal velocities as they transit constructive and destructive regions respectively. Despite appearances, this behaviour can be fitted with a classical relativistic framework.

\section{Theory}

\noindent We describe here, for the first time, the theoretical protocol for experimentally measuring the relativistic Bohmian velocity predicted by Foo et. al \cite{foo2022relativistic}. 
In the head-on interference of the counter-propagating wavefunctions---$\psi_R(x,t)$ and $\psi_L(x,t)$ in Fig.~\ref{schematic_setup}(a)---of a single photon in a Michelson-Sagnac interferometer along $x$ direction, the relativistic Bohmian velocity field is given by,
\begin{equation}
    \label{eqn_Foo_weak}
    v(x,t) = \frac{\langle \hat{k}_x \rangle ^{w}}{\langle \hat{H} \rangle ^{w}}.
\end{equation}
Here, $\hat{k}_x$ is the momentum operator in the $x$ direction, $\hat{H}$ is the Hamiltonian operator, and $\langle \rangle^w$ denotes the weak value \cite{aharonov1988result, duck1989sense}. For an observable given by the operator $\hat{A}$, its weak value $\langle \hat{A} \rangle^w$ is the average value of $\hat{A}$ obtained by weakly measuring $\hat{A}$ on an ensemble of particles, each prepared in a pre-selected state $\ket{\psi_{pre}}$ and post-selected on the state $\ket{\psi_{post}}$. Considering only the real part, the weak value $\langle \hat{A} \rangle^w$ is defined as follows:
\begin{equation}
    \langle \hat{A} \rangle^w = \Re{\frac{{\langle \psi_{pre}|\hat{A}|\psi_{post} \rangle}}{\langle \psi_{pre}| \psi_{post} \rangle}}.
\end{equation}
The superposition state of the photon resulting from the interference of a right-moving state $\ket{\psi_R}$ and a left-moving state $\ket{\psi_L}$ can be written as
\begin{equation}
    \label{eqn_superposition}
    \ket{\psi} = \sqrt{1-\eta} \ket{\psi_R} + \sqrt{\eta} \ket{\psi_L},
\end{equation}
where $\eta \in [0,1]$ parametrises the amplitudes of the states. We perform the analysis in natural units ($c=\hbar = 1$) and invoke the optical approximation where the spread of the frequency of wavefunctions is much smaller than their mean values. Given this approximation, the right and left propagating states $\ket{\psi_R}$ and $\ket{\psi_L}$ are the approximate eigenstates of both the operators $\hat{H}$ and $\hat{k}_x$. In the static laboratory frame of reference relevant to the experiment, both the eigenstates  $\ket{\psi_R}$ and $\ket{\psi_L}$ of the Hamiltonian operator are degenerate, having the same eigenvalue $\omega$. Thus, the weak value of energy can be written as follows:
\begin{equation}
\begin{split}
\label{eqn_weak_energy}
    \langle \hat{H} \rangle ^{w} &= \Re{\frac{{\langle x|\hat{H}|\psi \rangle}}{\langle x| \psi \rangle}} \\ &= \Re{\frac{\omega {\langle x|\psi \rangle}}{\langle x| \psi \rangle}} \\ &= \omega.
\end{split}
\end{equation}
The momentum operator acts on the right-moving and left-moving eigenstates $\ket{\psi_R}$ and $\ket{\psi_L}$ to yield the eigenvalues $k_x$ and $-k_x$ respectively. Thus, the momentum weak value takes the form:
\begin{equation}
\begin{split}
\label{eqn_weak_momentum}
    \langle \hat{k}_x \rangle ^{w} &= \Re{\frac{{\langle x|\hat{k}_x|\psi \rangle}}{\langle x| \psi \rangle}} \\  &= \Re{\frac{{\langle x|\hat{k}_x|  (\sqrt{1-\eta}\ket{\psi_R} + \sqrt{\eta}\ket{\psi_L}) \rangle }}{\langle x| \psi \rangle}} \\
    &= k_x \frac{\Re{\sqrt{1-\eta}\langle x | \psi_R \rangle - \sqrt{\eta}\langle x | \psi_L \rangle} }{\Re{\sqrt{1-\eta}\langle x | \psi_R \rangle + \sqrt{\eta}\langle x | \psi_L \rangle}}.
\end{split}    
\end{equation}
Up to a global phase, the complex probability amplitudes associated with the interferometer paths can be expressed as $\langle x | \psi_R \rangle = \abs{\langle x | \psi_R \rangle}$ and $\langle x | \psi_L \rangle = e^{i \phi} \abs{\langle x | \psi_L \rangle}$, where $\phi$ is the relative phase between the two paths. This relative phase $\phi$ is a function of $x$ and can be written as $\phi(x) = k_x x + \phi_0$, where $\phi_0$ is a constant phase offset. In our experiment, we use a continuous wave (CW) laser which makes $\abs{\langle x | \psi_R \rangle} = \abs{\langle x | \psi_L \rangle}$. Using a CW laser also makes the resulting relativistic Bohmian velocity independent of time. With the dispersion relation in natural units being $\omega = k_x$, the velocity in equation \ref{eqn_Foo_weak} takes the following form:
\begin{equation}
\label{eqn_weakRelBohm_vel_static_frame}
    \begin{split}
    v(x) &= \frac{\Re{\sqrt{1-\eta} - e^{i \phi} \sqrt{\eta}}}{\Re{\sqrt{1-\eta} + e^{i \phi} \sqrt{\eta}}} \\
    &= \frac{1 - 2 \eta}{1 + 2 \cos{\phi} \sqrt{\eta(1-\eta)}}.
    \end{split}
\end{equation}
Here, we multiplied both the numerator and denominator by $(\sqrt{1-\eta} + e^{-i \phi} \sqrt{\eta})$ and took the real part. 
We infer from equations \ref{eqn_weak_momentum} and \ref{eqn_weakRelBohm_vel_static_frame} that measuring the relativistic Bohmian velocity of the photon in our experiment is directly associated with measuring the photon's weak value of momentum $\langle \hat{k}_x \rangle ^{w}$ in the propagation direction.
Using the definition of the relativistic Bohmian velocity according to Foo et. al. \cite{foo2022relativistic} as the ratio of the Klein-Gordon probability current $j_x$ and the Klein-Gordon probability density $\rho$, we identify
\begin{align}
    \label{eqn_weak_model_1d_prob}
    \rho &\propto (1 + 2 \cos{\phi} \sqrt{\eta(1-\eta)}) \\
    \label{eqn_weak_model_1d_probcurr}
    j_x &\propto (1 - 2 \eta). 
\end{align}
In the standard interpretation, the probability current $j_x$ is associated with the ``which-path" information in the interferometer, whereas the probability density is related to the intensity of the interference fringes. Balancing the trade-off between the visibility of the interference and the ``which-path" information in an interferometer \cite{PhysRevA.60.4285}, we present an experimental scheme to measure the Bohmian velocity in equation \ref{eqn_weakRelBohm_vel_static_frame}. We weakly measure the ``which-path" information whilst simultaneously preserve high-visibility interference fringes.  

\section{Experiment and Methods}

\begin{widetext}
\begin{figure*}[t]
    \centering
    \includegraphics[width=1.0\linewidth]{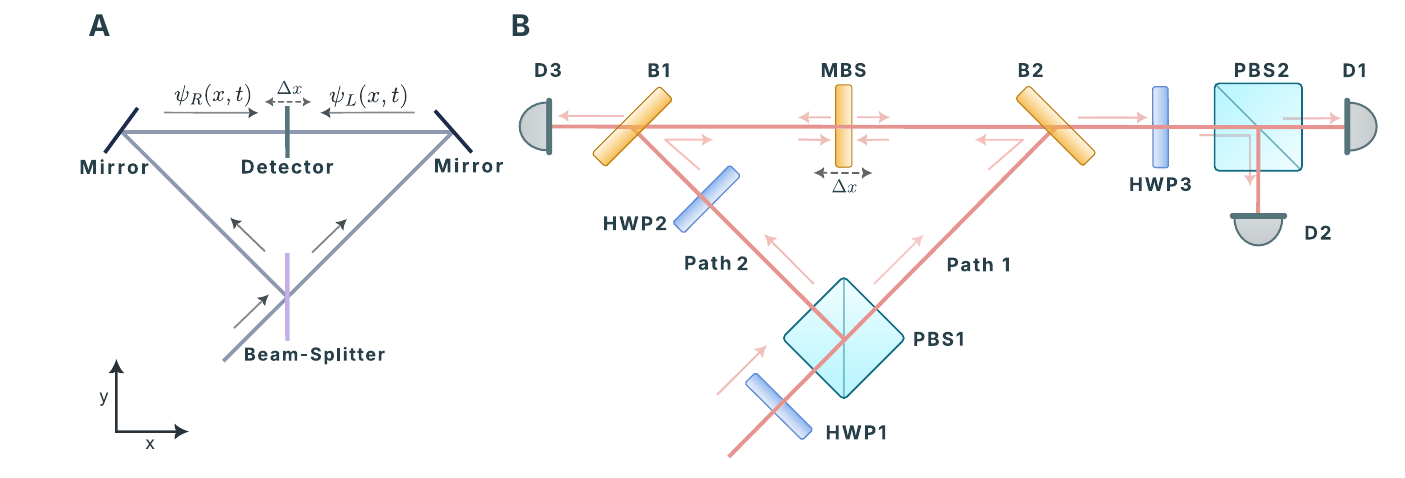}
    \caption{\textbf{Conceptual and experimental schematic.} \\
    \textbf{(A)} Conceptual setup of the Michelson-Sagnac interferometer consisting of the beam-splitter in purple and the mirrors in black. The detector in dark grey measures the weak value of momentum $\langle \hat{k}_x \rangle ^{w}$, at different positions $x$, allowing the determination of the relativistic Bohmian trajectories of a single photon which is in a superposition of a right-moving wavefunction $\psi_R(x,t)$ and a left-moving wavefunction $\psi_L(x,t)$. \textbf{(B)} The schematic of the Michelson-Sagnac interferometer used to implement the conceptual setup. It consists of polarising beam-splitters PBS1 and PBS2 in cyan, half-waveplates HWP1, HWP2, and HWP3 in blue, plate beam-splitters B1, B2, and MBS (measurement beam-splitter) in yellow, and the single photon detectors D1, D2, and D3 in grey. The MBS is translated along $x$ direction to probe both the interference between the two wavefunctions and the momentum information of the photon at different $x$ positions, thus probing the 1-D velocity. The propagation of light through the interferometer is shown by light red lines with arrows, starting from input to HWP1. 
    }
    \label{schematic_setup}
\end{figure*}
\end{widetext}

\noindent Operationally, our experiment requires the measurement of the momentum of the photon as well as probing the interference fringes at the position of the detector in the conceptual setup in Fig. \ref{schematic_setup}(a) to reconstruct the relativistic Bohmian velocity as in equation \ref{eqn_weakRelBohm_vel_static_frame}. The schematic in Fig. \ref{schematic_setup}(b) depicts the implementation of the conceptual setup in Fig. \ref{schematic_setup}(a).
A single photon is injected into the interferometer through a variable beam-splitter of transmissivity $\eta$ implemented using a half-wave plate HWP1 and a polarising beam-splitter PBS1. This variable beam-splitter determines the probability amplitude of the photon wavefunctions to propagate along path 1 or path 2 of the interferometer. The experiment probes both the interference between the two wavefunctions and the momentum information of the photon at different positions along $x$. This position $x$ is determined by the position of a 50:50 beam-splitter, referred to as the measurement beam-splitter (MBS) in Fig. \ref{schematic_setup}(b). 
The polarisation degree of freedom of the photon is used as the pointer to perform the weak-value measurement of the photon's momentum. We introduce a small deviation of an angle $\theta$ of the fast axis of the half-waveplate HWP2 in path 2 away from $45^\circ$ w.r.t. its optic axis. At $45^\circ$, HWP2 transforms the vertically polarised light emerging from the reflected output of the PBS1 to horizontally polarised light. The additional small rotation $\theta$ weakly encodes the photon's momentum information. Finally, photon counts are measured in the linear-polarisation basis using a combination of the half-waveplate HWP3 and polarising beam-splitter PBS2. 
The angle of the fast axis of HWP3 is chosen to be offset by $\theta'$ relative to $22.5^\circ$ w.r.t. its optic axis. At $22.5^\circ$, HWP3 converts the horizontally polarised light to diagonally polarised light. We relate the transformation of the polarisation state by HWP2 with a beam-splitter transformation with reflectivity $\epsilon$. We associate the variable beam-splitter comprised of HWP3 and PBS2 in the experiment with a beam-splitter of reflectivity $\epsilon'$ (see description of a photonic circuit equivalent to the experimental schematic of Fig.~\ref{schematic_setup}(b) in the supplementary). To achieve a balance of interference and ``which-path" information in the experiment, $\epsilon$ and $\epsilon'$ are related by the following parametrisation of $\delta$:
\begin{align}
\label{eqn_epsilon_epsilon'_delta}
    \epsilon &= 1 - \delta \\
    \epsilon' &= \frac{1}{2}\Big(1 - \sqrt{\delta}\Big).
\end{align}
The parameter $\delta$ is expressed in terms of the physical rotation $\theta$ of the half-waveplate HWP2 of the schematic in Fig.~\ref{schematic_setup}(b) as $\delta = \sin^2{2 \theta}$. Ensuring the relation between $\epsilon$ and $\epsilon'$ as in equation \ref{eqn_epsilon_epsilon'_delta} implies, for small angle $\theta$, the relation between $\theta$ and $\theta'$ comes out to be (see supplementary material for derivation):
\begin{equation}
    \label{eqn_relation_theta_theta'}
    \theta = 2 \theta'.
\end{equation}

\noindent With $\hat{N_1}$ and $\hat{N_2}$ denoting the number operators corresponding to the counts in the detectors $D1$ and $D2$ respectively such that $\langle \hat{N}_1 \rangle = N_1$ and $\langle \hat{N}_2 \rangle = N_2$, and $\phi$ being the relative phase between the paths in the interferometer, the average of the sum and difference of the photon counts are given by: 
\begin{align}
    \label{eqn_sum_counts_photonic}
    \langle \hat{N_1} + \hat{N_2} \rangle &= \frac{1}{2}\big(1 + 2\cos{\phi} \sqrt{\eta (1-\eta)}\sqrt{\epsilon}\big)  \\
    \label{eqn_diff_counts_photonic}
    \langle \hat{N_1} - \hat{N_2} \rangle &= \frac{1}{2} \big(1 - 2 \eta \big) \sqrt{\delta}.
\end{align}

\noindent A full derivation of these equations by Heisenberg evolution of the mode operators is provided in the supplementary material.
For small $\delta$, implying $\epsilon \to 1$ (see equation \ref{eqn_epsilon_epsilon'_delta}), 
we arrive at the following equation to experimentally reconstruct the relativistic Bohmian velocity $v(x)$ as in equation \ref{eqn_weakRelBohm_vel_static_frame}:
\begin{equation}
    \label{eqn_weakRelBohm_vel_counts}
    v(x) = \frac{1}{\sqrt{\delta}} \frac{\langle \hat{N_1} - \hat{N_2} \rangle}{\langle \hat{N_1} + \hat{N_2} \rangle}.
\end{equation}

\noindent We probe the Bohmian trajectories using a weak coherent state provided by a heavily attenuated CW laser operating at 925 $nm$. Single photons are detected using superconducting nanowire single photon detectors (SNSPDs): since the experiment is based on the first order interference and the photon number statistics has no role to play, an attenuated laser is suitable \cite{barnett2022single}. 
For a single photon, the propagation of light is to be understood in terms of probability amplitudes in the two paths of the interferometer. 
We carry out the experiment by fixing the offset $\theta$ of HWP2 on path 2 of the interferometer at $\theta {=} 4^{\circ}$. The offset of HWP3 outside the interferometer is set to $2^{\circ}$ according to equation \ref{eqn_relation_theta_theta'}. 
The MBS is mounted on a motorised translation stage and scanned over a range of 1 $\mu m$ in 20 $nm$ increments. 
At each position of the MBS, the counts are recorded for 60 $s$ by three detection channels D1, D2, and D3. Since the reconstruction of relativistic Bohmian trajectory is based on a single-photon theory, we ensure the experiment is in low photon regime considering only photon detections where exactly only one of D1, D2, or D3 fires in a 1 $ns$ interval. In the experiment, such occurrence of simultaneous counts in 1 $ns$ interval is below $10^{-3}$. Although the detection channel D3 does not take part in the reconstruction of the relativistic Bohmian velocity, it plays a crucial role in ensuring the experiment is in sufficiently low photon regime.
The Bohmian velocity $v(x)$ as a function of the position is reconstructed from the photon counts as described in equation \ref{eqn_weakRelBohm_vel_counts}.  \\

\noindent The effective detection efficiency of the three detection channels depend on the respective coupling efficiencies from free space to fiber, and different efficiencies of the single-photon detectors. Since the photon counts ($N_1$ and $N_2$) from two detection channels D1 and D2 respectively are considered for reconstructing the relativistic Bohmian velocity, we define the effective detection efficiency $\kappa_{eff}$ as 
\begin{equation}
\label{eqn_det_efficiency}
    \kappa_{eff} = \frac{\kappa_{c2}}{\kappa_{c1}} \times \frac{\kappa_{d2}}{\kappa_{d1}}.
\end{equation}
Here, $\kappa_{c1}$ and $\kappa_{c2}$ are the coupling efficiencies from free space to fiber in the detection channels D1 and D2 respectively. The SNSPD detector ports of D1 and D2 have efficiencies $\kappa_{d1}$ and $\kappa_{d2}$ respectively.
In the experiment, the effective detection efficiency $\kappa_{eff}$ was 0.62. The relativistic Bohmian velocity $v_{expt}(x)$ is reconstructed from the experiment in terms of the average of the sum and difference of photon counts ($N_1$ and $N_2$), the angle offset $\theta$ of the half-wave plate HWP2, and the effective detection efficiency $\kappa_{eff}$ as (see equation \ref{eqn_weakRelBohm_vel_counts}):
\begin{equation}
    \label{eqn_vel_expt}
    v_{expt}(x) = \frac{\kappa_{eff}N_1(x) - N_2(x)}{\sin{(2 \theta)} (\kappa_{eff}N_1(x) + N_2(x))}.
\end{equation}
The theoretical relativistic Bohmian velocity $v_{theo}(x)$ is calculated from
average photon counts as (see equations \ref{eqn_sum_counts_photonic} and \ref{eqn_diff_counts_photonic}): 
\begin{equation}
    \label{eqn_vel_theo}
    v_{theo}(x) = \frac{1 - 2 \eta}{1 + 2 \cos{[\phi(x)]} \sqrt{\eta(1-\eta)}\sqrt{\epsilon}}.
\end{equation}   
The parameter $\eta$ is fixed by a particular splitting ratio in the interferometer of the schematic in Fig.~\ref{schematic_setup}(b). This splitting ratio refers to the ratio of probability amplitudes of the wavefunctions of the the single photon in the different paths 1 and 2 of the interferometer. 
The parameter $\epsilon$, where $\epsilon {=} 1 {-} \sin^2(2\theta)$, is fixed at ${\approx} 0.14$ for $\theta {=} 4^{\circ}$. The probability density varies only due to variation in $\phi$. The reconstructed velocity along $x$ changes with different MBS positions ($x$ positions) according to the variation in probability density while the probability current remains constant. \\ 

\noindent We represent the velocity $v(x)$ as a function of $x$---the detector in the conceptual setup and the MBS in the schematic in Fig.~\ref{schematic_setup}---in terms of velocity field $v(x,t)$ and plot the trajectories with the probability density in the background for better visualisation.
To obtain the Bohmian velocity field $v(x,t)$, we incorporate the temporal dimension in the reconstructed velocity $v(x)$. This is achieved by approximating a CW wavepacket as an infinitely extended pulse in time. Thus, the probability density obtained by the interference of the wavefunctions approaching head-on, can be assumed to be constant in the temporal dimension. The probability current is constant in both $x$ and $t$ and depends only on splitting ratio of the intensity of light in the two paths of the interferometer. This feature of constant probability density and current in time enables to plot trajectories corresponding to the velocity field $v(x,t)$ by stacking the velocity vectors in and joining them (see supplementary material) in a spacetime ($x$-$ct$) plot. 

\begin{widetext}
\begin{figure*}[t]
    \centering
    \includegraphics[width=1.0\linewidth]{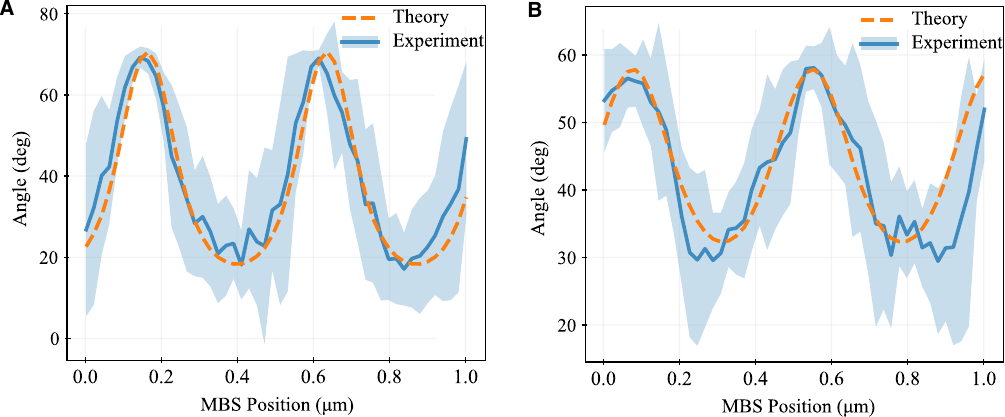}
    \caption{\textbf{Velocity-vector-angles as a function of MBS position.} \\
    \textbf{(A)} and \textbf{(B)} illustrates experimental velocity-vector-angles compared with the theory as a function of MBS position for $20$-$80$ and $5$-$95$ intensity splitting ratios in the interferometer, respectively. The theoretical plot is shown in dashed orange lines and the experimental data are shown in blue solid lines with blue shaded region of uncertainty by $\pm$ 1 standard deviation.}
    \label{Inter_angle}
\end{figure*}
\end{widetext}

\begin{widetext}
\begin{figure*}[t]
    \centering
    \includegraphics[width=1.0\linewidth]{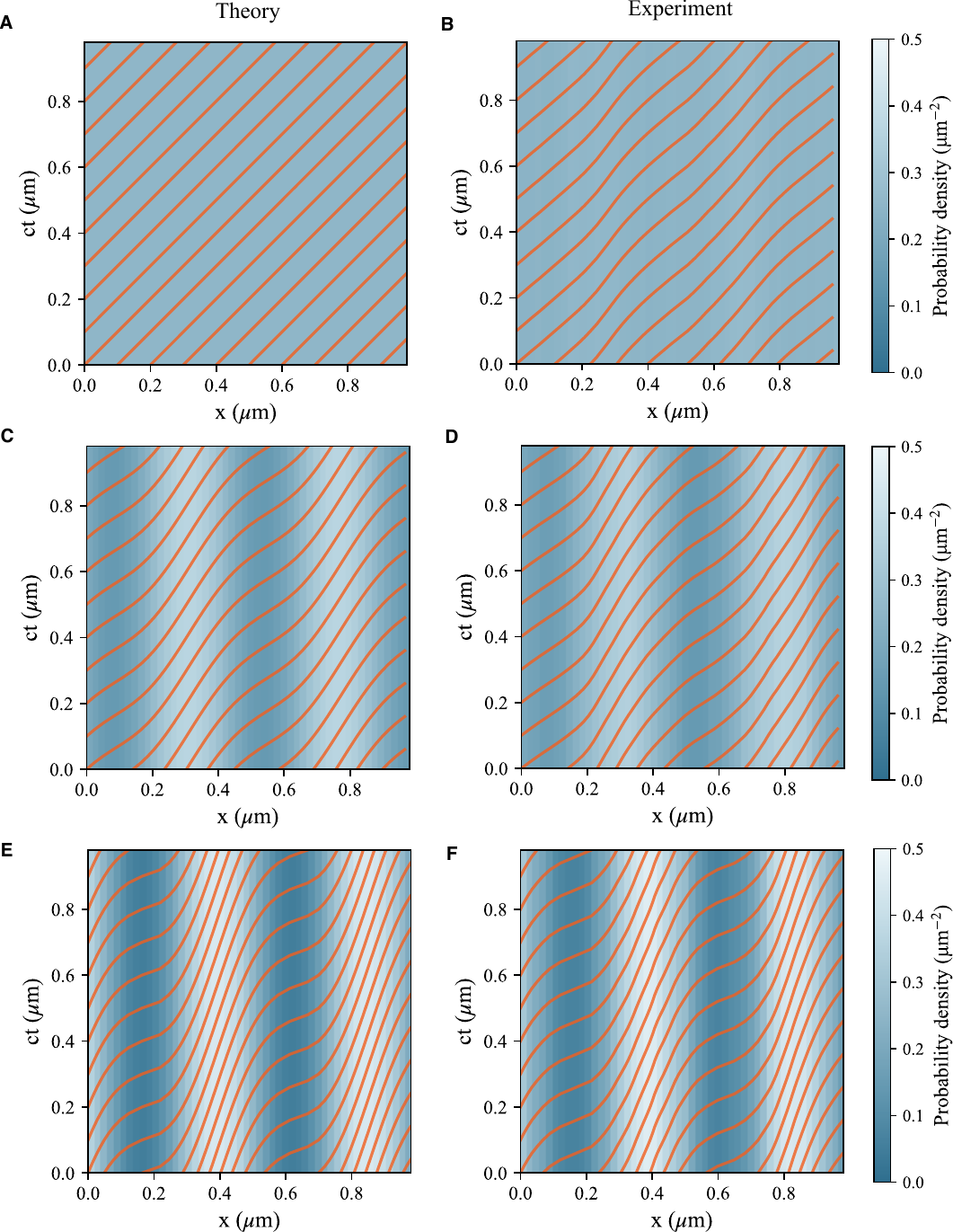}
    \caption{\textbf{Theoretical and experimental relativistic Bohmian trajectories overlaid on the probability densities.} \\
    \textbf{(A)}, \textbf{(C)}, and \textbf{(E)} show the theoretical trajectories in dark orange on the theoretical probability densities. \textbf{(B)}, \textbf{(D)}, and \textbf{(F)} illustrate the experimental trajectories in dark orange on the experimental probability densities. The common colourmap for the probability densities are shown to the right of each row.}
    \label{traj}
\end{figure*}
\end{widetext}

\section{Results}

\noindent We show the Bohmian-velocity-vector-angle $\chi {=} \tan^{-1}[v(x)]$ along with its uncertainty (shaded region) as a function of the MBS position in Fig.~\ref{Inter_angle}(a) and Fig.~\ref{Inter_angle}(b) for the cases of $20$--$80$ and $5$--$95$ splitting ratios, respectively, in the interferometer. The trajectories overlaid on the probability density are illustrated in Fig. \ref{traj} (theory to the left and experiment to the right). Fig. \ref{traj}(a) and (b) shows the straight line trajectories at $45^{\circ}$ and the constant probability density for the case of no interference for a $0$--$100$ splitting ratio in the interferometer. This illustrates that the relativistic Bohmian velocity is the velocity of light in the absence of interference. 
The experimental trajectories not being exactly at $45^{\circ}$ is attributed to errors and small amount of residual interference (see related discussion in the supplementary material). As we introduce interference, the photon trajectories begin to curve as shown in Fig.~\ref{traj}(c) and Fig.~\ref{traj}(d) for $5$--$95$ splitting ratio and in Fig.~\ref{traj}(e) and (f) for $20$--$80$ splitting ratio. These curved trajectories exhibit a feature of the underlying physics of the experiment rather than an artifact of the experimental noise, as supported by the theoretical trajectory plots of Fig.~\ref{traj}(c) and Fig.~\ref{traj}(e), and plots of the angle as a function of MBS position in Fig~\ref{Inter_angle}. The Bohmian velocity-vector-angle $\chi$ having a magnitude greater and lesser than $45^{\circ}$ signifies superluminal and subluminal velocities respectively. The trajectories bunch together and propagate at subluminal speeds in regions of high probability density. In contrast, they diverge from one another and exhibit superluminal motion in regions of low probability density. From the angle plots in Fig.~\ref{Inter_angle} and the trajectory plots in Fig.~\ref{traj}(c), Fig.~\ref{traj}(d), Fig.~\ref{traj}(e), and Fig.~\ref{traj}(f), it is evident that the contrast between the subluminal and superluminal velocity-vector-angle increases with higher interference visibility in $20$--$80$ splitting ratio than $5$--$95$ splitting ratio. This behaviour is consistent with the fact that the relativistic Bohmian velocity in equation \ref{eqn_weakRelBohm_vel_counts} changes along with MBS position as the average sum of the counts change whereas the average difference of counts remain constant. We note that, approaching further towards the $50$--$50$ splitting ratio makes the behaviour of the trajectories more sensitive to experimental errors (see supplementary material for discussion and examples). If the splitting ratios are reversed, e.g. $80$--$20$, the photon trajectories follow the opposite path through the interferometer (see supplementary material for examples). \\

\section{Discussion and Conclusion}

\noindent The consistency of the Bohmian velocity with relativity is demonstrated by expressing the velocity $v(x,t) {=} j_x/\rho$ as the ratio of the Klein-Gordon probability current $j_x$ and Klein-Gordon probability density $\rho$ \cite{foo2022relativistic} (see equations \ref{eqn_weak_model_1d_prob} and \ref{eqn_weak_model_1d_probcurr}). In the (1+1)-D  spacetime, $j_x$ and $\rho$ are the components of the Klein-Gordon charge-current two vector, and hence, transform covariantly under Lorentz transformation \cite{bromley2013relativistic}. The continuity equation ($\frac{\partial \rho}{\partial t} {+} \frac{\partial j_x}{\partial x} {=} 0$) ensures that the density of trajectories corresponds to the probability density, in accordance with the Born rule. We note that, the Bohmian velocity is defined as a coordinate velocity $v(x,t) {=} j_x/\rho$ as opposed to a local velocity \cite{foo2022relativistic, hartle2021gravity}. In relativity, coordinate velocities of light can be subluminal or superluminal given the presence of spacetime curvature \cite{hartle2021gravity}. Here, curvature replaces the Bohmian quantum potential of non-relativistic theories. \\

\noindent It is interesting to analyse the results of the experiment from the different perspectives of the Copenhagen and the Bohmian interpretations. Standard quantum mechanics describes the fringes resulting from interference of single-photon probability amplitudes, with the position of the photon within the interferometer undefined unless a measurement is made. The Copenhagen interpretation has indeterminism as a main feature. The Bohmian picture assigns the photon a definite path (1 or 2) through the interferometer with the its velocity fluctuating between greater or lesser than the nominal velocity of light $c$ as it moves through the region of fringes. This change in velocity can be explained by spacetime curvature. Similar to the Bohmian potential in non-relativistic theories, the strength of the curvature depends on the wavefunction but its source is not specified by the theory. \\

\noindent To conclude, we experimentally measured the apparently surprising predictions of sub- and superluminal relativistic Bohmian photon trajectories by Foo et. al.~\cite{foo2022relativistic}. The head-on collision geometry of the interference of the photon's wavefunctions and the weak measurement of the photon's momentum performed in its propagation direction elegantly capture the full relativistic nature of the measurement. Our work paves the way for exploring future experiments to measure relativistic Bohmian trajectories for entangled photons, a pathway to explore nonlocality in the relativistic regime. Another intriguing future scope is to investigate the experiment for boosted reference frames, leading to measurement of backwards-in-time trajectories. \\

\noindent We note that, an experiment taking a different approach to realising the relativistic proposal has been recently reported \cite{wang2025directlyobservingrelativisticbohmian} in a double slit interferometer setup.

\section{Acknowledgments}
\noindent We thank Joshua Foo for helpful theoretical discussions. We are grateful to Farzad Ghafari and Giulio Gualandi for insightful discussions on the experiment. We acknowledge the use of a large language model in constructing and optimising the codes for acquiring and post-processing the experimental data.

\section{Code Availability}
\noindent The codes acquiring the experimental data, processing the acquired data, and generating the plots are found in \cite{Rel_Bohm_expt_codes}.


\bibliographystyle{apsrev4-2}
\bibliography{references}

\end{document}